# Exploration of amorphous $V_2O_5$ as cathode for magnesium batteries

Vijay Choyal, Debsundar Dey, Gopalakrishnan Sai Gautam*

Department of Materials Engineering, Indian Institute of Science, Bengaluru, 560012, India

*Email: saigautamg@iisc.ac.in

**Abstract**

Development of energy storage technologies that can exhibit higher energy densities, better safety, and lower supply-chain constraints than the current state-of-the-art Li-ion batteries is crucial for our transition into sustainable energy use. In this context, Mg batteries offer a promising pathway to design energy storage systems with superior volumetric energy densities than Li-ion but require the development of positive electrodes (cathodes) that can exhibit high energy densities at a reasonable power performance. Notably, amorphous materials that lack long range order (LRO) can exhibit 'flatter' potential energy surfaces than crystalline frameworks, possibly resulting in faster $Mg^{2+}$ motion. Here, we use a combination of density functional theory (DFT), ab initio molecular dynamics (AIMD), and machine learned interatomic potential (MLIP) based calculations to explore amorphous $V_2O_5$ as a potential cathode for Mg batteries. Using a DFT/AIMD-generated dataset, we train and validate moment tensor potentials that can accurately model amorphous $V_2O_5$ and $MgV_2O_5$, as verified by the calculated radial distribution functions and LRO. Due to the amorphization of $V_2O_5$, we observe a 10-14% drop in the average Mg intercalation voltage – but the voltage remains higher than sulfide and selenide Mg cathodes. Importantly, we find a ~seven orders of magnitude higher $Mg^{2+}$ diffusivity in amorphous $MgV_2O_5$ than its crystalline version, which is directly attributable to the amorphization of the structure. Additionally, the $Mg^{2+}$ diffusivity in amorphous $MgV_2O_5$ is higher by ~five orders of magnitude compared to the thiospinel $Mg_xTi_2S_4$ cathode. Also, we note the $Mg^{2+}$ motion in the amorphous structure is significantly cross-correlated at low temperatures, with the correlation decreasing with increase in temperature. Thus, our work highlights the potential of amorphous $V_2O_5$ as a cathode that can exhibit both high energy and power densities, resulting in the practical deployment of Mg batteries. Moreover, amorphization of oxides provides an important design handle that can enable the development of other Mg cathodes. Finally, our theoretical framework of employing DFT/AIMD-trained MLIPs for performing large-scale and long-range simulations can be extended to study other amorphous systems with applications in energy storage and beyond.

# 1. Introduction

Energy storage technologies are essential for driving humanity's transition to a sustainable future, by facilitating electric vehicles, grid-scale storage, and portable electronics.[1–3] While lithium-ion batteries (LIBs) have long dominated the energy storage landscape due to their high energy density, power performance, and long cycle life, LIBs are increasingly approaching their fundamental limitations,[4–6] which include resource scarcity and supply-chain constraints, safety concerns, and performance degradation over prolonged use.[7] As the demand for more efficient and sustainable energy storage solutions grows, magnesium batteries have emerged as a promising alternative,[8–11] since it enables the use of a metallic anode that results in higher volumetric energy density,[8,12] the lower tendency of Mg metal to form dendrites that enhances safety[13,14] and Mg being a fairly abundant element on the earth's crust that can reduce supply-chain constraints. Despite these advantages, the widespread adoption of Mg batteries is hindered by the poor diffusion of $Mg^{2+}$ in crystalline positive electrode (cathode) materials, especially high energy density oxide chemistries, and the susceptibility to conversion reactions,[15] resulting in poor rate performance and cycle life that limits practical deployment.[11,16–19] Thus, designing new cathodes that can exhibit high energy densities and reasonable power densities is critical for the advancement of Mg batteries, which is the focus of this work.

State-of-the-art (SOTA) Mg batteries utilize low-voltage (and hence low energy density) sulfide cathodes, such as Chevrel-$Mg_xMo_6S_8$[20] and thiospinel-$Mg_xTi_2S_4$,[21] which offer structural stability, reasonable rate performance, and reversible Mg intercalation. In case of oxides, several materials that have shown an ability to intercalate $Li^+$ have been explored as Mg cathodes in an attempt to improve the energy density of Mg batteries. For example, $V_2O_5$ is one of the first oxides to be explored as a potential cathode for Mg batteries,[22–26] with experiments often reporting limited Mg intercalation (and hence limited capacity) due to poor kinetics[27] and capacity fade with cycling. Spinel-$Mn_2O_4$ has been shown to electrochemically intercalate Mg ions at a voltage of ~2.9 V, but its practical performance is limited by cation inversion in the structure that blocks Mg diffusion pathways.[28] Spinel-$Cr_2O_4$ offers a higher theoretical voltage (~3.6 V) and is not as prone to cation inversion compared to $Mn_2O_4$,[29] but suffers from synthesis bottlenecks and electrolyte instability at high voltages. Mixed transition metal spinel oxides, such as Cr-V and Cr-Mn, have reported limited success in intercalating Mg in the bulk, with capacity fade still an ongoing challenge.[30,31] Theoretical calculations have predicted facile Mg transport in post-spinel oxides,[32] but experimental validation of such predictions are not available so far. Although several studies have reported improved Mg cathode performance with the introduction of water into the electrolyte,[8] with solvent intercalation often attributed to aid in Mg diffusion in such cases,[33,34] the capacity and rate performance improvements with water addition can also be caused by proton intercalation.[35] Thus, enhancing the rate performance in oxide-based chemistries, via improvements in Mg mobility in the bulk, is critical in the design of cathodes that can deliver higher energy densities compared to the SOTA Mg cathodes.

One of the reasons attributed for poor Mg diffusion (compared to Li) in oxide lattices is the stronger electrostatics of the $Mg^{2+}$-$O^{2-}$ bonds (versus $Li^+$-$O^{2-}$) resulting in the need for larger distortions within the structure as $Mg^{2+}$ migrates within the lattice.[36] In other words, the stable site(s) that Mg sits in a given lattice is often 'deep' within the potential energy surface (PES), which in turn creates a large migration barrier ($E_a$) that $Mg^{2+}$ must cross for macroscopic diffusion. Indeed, the insertion of $Mg^{2+}$ into anatase-$TiO_2$ has been shown to induce cooperative lattice distortion that deepen the energy of the stable sites that Mg occupies instead of perturbing the transition state's energy during migration, resulting in an increase in $E_a$ from 537 to 1500 meV with Mg insertion.[37] One strategy to reduce $E_a$, as proposed by Rong et al.,[17] is to use structures where $Mg^{2+}$ occupy 'unpreferred' (i.e., non-octahedral) coordination environments, which can increase the energy of the stable sites thereby 'flattening' the PES and reducing $E_a$. While the strategy of Rong et al. has shown limited success, such as the identification of spinel Mg-ionic conductors[38,39] and the use of spinel-oxides as Mg cathodes,[18,28,40,41] the mobility enhancements in oxides remain insufficient to surpass the performance of SOTA Mg cathodes.

Another approach to achieve a flatter PES is to reduce the long range order (LRO) of oxides, via amorphization for example, which can reduce the depth of the stable sites that $Mg^{2+}$ occupy and facilitate Mg mobility. Note that diffusion is known to be significantly faster in the highly defective regions (such as grain boundaries) rather than the perfect crystalline regions of a microstructure, especially in systems where the diffusivity in the bulk is 'low' (~$10^{-14}$ $cm^2$/s or below at 300 K), as in the case of metals.[42] In grain boundaries, as with amorphous solids, the LRO of the lattice is disrupted, resulting in an increase in the energy of the 'stable' sites that atoms occupy and a flatter PES. Therefore, 3$d$ transition metal containing redox-active oxide chemistries that can exhibit an amorphous structure may improve Mg mobility in the 'bulk' significantly and can function as potential cathodes in Mg batteries. However, such amorphous structures are likely to be metastable and hence reduce the Mg (de)intercalation voltages. Nevertheless, it is worthwhile to explore if amorphous oxides can be used as possible cathodes for Mg batteries. Given that $V_2O_5$ has been explored as a cathode before for Mg batteries, in both its bulk orthorhombic and nanocrystalline xerogel forms,[11,26,27,33,34] and the ability of $V_2O_5$ to become amorphous[43] and intercalate ions,[44,45] we choose $V_2O_5$ as a possible amorphous oxide cathode for Mg batteries in this work.

Modelling amorphous structures computationally is non-trivial, since structural models need to be 'large' enough to ensure LRO is broken over sufficiently long distances and the sampling of the ionic dynamics needs to be 'long' enough so that the transport properties can be accurately estimated. While density functional theory (DFT) and ab initio molecular dynamics (AIMD) calculations enable accurate predictions of material properties, both techniques have severe limitations on the system sizes (~few 100s) and time scales (~100 ps) that can be accessed.[46] Notably, machine learned interatomic potentials (MLIPs) that are trained on small-scale DFT/AIMD-generated datasets, can provide both 'quick' and 'accurate' estimates of energies and atomic forces, enabling classical molecular dynamics (MD) simulations that can sample large length and long time scales.[47–54] Typically, MLIPs learn the influence of different 'local' coordination environments on an atom of interest, from the DFT/AIMD-generated dataset, to predict the energy and force on the atom.[55] Here, we choose the moment tensor potential (MTP[56,57]) framework, since its accuracy and computational speed has been showcased in several studies, such as predicting ionic diffusivities in solid electrolytes,[58] modelling ionic transport across interfaces,[59] and describing the PES of multi-component systems.[60] MTP models the local coordination environment (within a cut-off radius) around an atom of interest via contracted moment tensors, which consist of radial distribution functions to describe distances and outer products of position vectors of neighbouring atoms to describe angular interactions.[57] Additionally, MTP incorporates an active learning framework[61] that can be used to validate and refine a pre-trained MTP when employed in larger MD simulations.

In this study, we combine DFT/AIMD and MTP-based MD simulations to explore the utility of amorphous-$V_2O_5$ as a potential cathode for Mg batteries. To construct the MTP, we generate an AIMD-simulated dataset of $V_2O_5$ and $MgV_2O_5$ across different temperatures, using the melt-quench technique, resulting in an overall dataset of 3156 configurations. Upon training the MTP on a train subset and optimizing its hyperparameters on the test subset, we fine-tune and validate the trained MTP using active learning, resulting in a final overall dataset of 3725 configurations. To ensure the amorphous nature of the structures generated using AIMD and MTP-MD, we examine the radial distribution functions (RDFs) and LRO at different temperatures in both $V_2O_5$ and $MgV_2O_5$. Subsequently, we quantify the impact of amorphization on Mg intercalation voltages and perform large-scale (2×4×6 supercell) and long-time (4 ns) MTP-based MD simulations to quantify Mg transport within amorphous $MgV_2O_5$. We observe a 10-14% drop in the average Mg intercalation voltage (versus Mg metal) due to amorphization of the $V_2O_5$ structure. Importantly, we find an increase in Mg diffusivity by ~seven orders of magnitude in amorphous-$MgV_2O_5$ compared to crystalline-$(Mg)V_2O_5$, corresponding to a low effective $E_a$ of 47 meV in the amorphous structure. Moreover, the Mg diffusivity in amorphous-$MgV_2O_5$ is higher than the SOTA $Mg_xTi_2S_4$ by ~five orders of magnitude at 300 K, while maintaining a higher average intercalation voltage than the SOTA. Also, we observe Mg motion to be highly cross-correlated at low temperatures with the motion becoming increasing random with increasing temperature. Thus, our work indicates the impact that amorphization can have on improving the Mg mobility in oxide lattices while ensuring that the energy density remains higher than SOTA sulphide and selenide cathodes. We are hopeful that our work will facilitate the practical deployment of Mg batteries by the design of optimized high-energy-density amorphized oxide cathodes. Finally, our theoretical framework is general and can be used to explore other promising amorphous compositions/structures, for Mg battery applications and beyond.

## 2. Methods

### 2.1. Workflow

**Figure 1** displays the workflow of our study, including data generation and the calculations employed, training of MTP, and active learning to further refine and validate the MTP for the $V_2O_5$ and $MgV_2O_5$ systems. After obtaining the initial structure of $V_2O_5$ from the inorganic crystal structure database (ICSD[62]), we performed melt-quench AIMD simulations to obtain amorphous structures of $V_2O_5$, which forms the computed dataset of $V_2O_5$. Subsequently, we identified possible sites that Mg can occupy within the amorphous structures using the TopographyAnalyzer[63] class of pymatgen,[64] and performed further AIMD calculations to create the $MgV_2O_5$ dataset. Combining the generated amorphous $V_2O_5$ and $MgV_2O_5$ datasets to create the overall dataset, we divided the overall dataset randomly into a 90:10 training:test split, and used the training subset to construct our MTPs. During training, we optimized the hyperparameters of MTP to minimize the root mean square error (RMSE) and the mean absolute error (MAE) with respect to AIMD-calculated energies and forces on the train set, with the optimized hyperparameters compiled in **Table S1** of the supporting information (SI).

Further, we tested our MTPs by performing MD simulations with active learning[61] to refine the training dataset and the validate MTP itself. After ensuring that our hyperparameter-optimized MTPs provided minimal breaks during the active learning step, we subsequently performed larger-size and longer-time MD simulations at different temperatures to obtain Mg intercalation voltages and diffusivities.

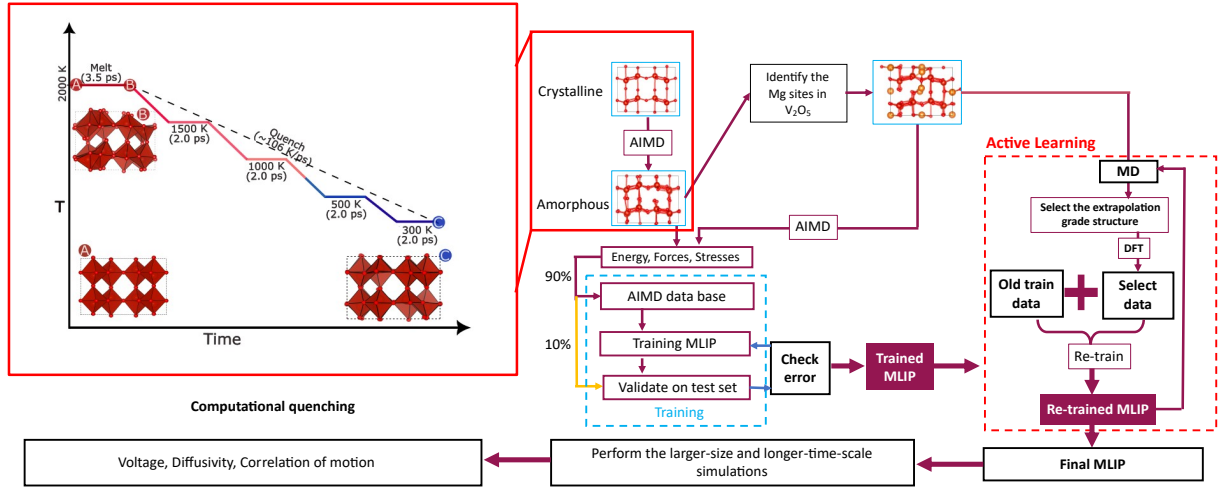

**Figure 1.** Workflow of dataset generation, DFT calculations, and MLIP training employed in this work. Train and test subsets are used for training and optimizing hyperparameters of the MLIP, respectively. The accuracy of the trained MLIP is validated by using active learning. Voltages and diffusivities are extracted from long-range and large-scale MD simulations performed using the optimized MLIP.

**2.2. Dataset generation and MTP construction**

We generated amorphous $V_2O_5$ structures via the well-known melt-quench simulation technique[65,66] using AIMD calculations. Specifically, we heated $V_2O_5$ to a temperature of 2000 K, which is beyond its melting point.[27,67] Subsequently, we quenched the molten $V_2O_5$ in steps, at an overall cooling rate of 106 K/ps. During quenching, we cooled and equilibrated the $V_2O_5$ structure for 2 ps at temperatures of 1500 K, 1000 K, and 500 K, before eventually quenching to 300 K. Subsequently, we extracted structures and equilibrated them at 1200 K, 900 K, and 600 K during the quenching process, to give us access to amorphous $V_2O_5$ structures at these intermediate temperatures. We tracked the radial distribution functions (RDFs) of V-V, V-O, and O-O bonds and the long range order (LRO) to verify the amorphous nature of the quenched structure at different temperatures (i.e., 1200 K, 900 K, and 600 K). Eventually, our AIMD simulations runs (and active learning) gave rise to 2518 configurations that constituted the $V_2O_5$ dataset, which was split into 2292 configurations for training and 226 for testing.

In order to generate the dataset of $MgV_2O_5$, we identified potential sites that Mg can occupy in the amorphous $V_2O_5$ structure at 300 K, and each of the three intermediate temperatures, i.e., 1200 K, 900 K, and 600 K. The number of possible sites Mg can occupy is typically higher than the stoichiometrically allowed number of occupied Mg sites to maintain a composition of $MgV_2O_5$. Hence, we enumerated symmetrically distinct occupations of Mg atoms among the possible sites that satisfied the $MgV_2O_5$ stoichiometry, ranked the different configurations based on their electrostatic energy (as calculated by the Ewald summation technique[68] and chose the lowest electrostatic energy structure for further AIMD simulations. Our choice of this workflow is to mimic a typical topotactic insertion process that occurs in intercalation electrodes.[69,70] The final amorphous $MgV_2O_5$ dataset (including active learning) generated contains 1207 configurations out of which 151 were used for testing. Note that we sampled one structure for every 50 time steps in our AIMD simulations of $V_2O_5$ and $MgV_2O_5$ at 300 K, 600 K, and 900 K, while we sampled one structure for every 20 time steps in our simulations at 1200 K, since the configurations at 1200 K are typically more disordered compared to the lower temperatures. Combining our $MgV_2O_5$ and $V_2O_5$ datasets, our overall dataset thus contains 3725 configurations, which we split into training (3348) and test (377) subsets.

A detailed description of the MTP framework can be found elsewhere.[57,60] Initially, we used the training set generated from AIMD calculations to train our MTP and obtained the optimal hyperparameters (**Table S1**) by minimizing errors in the test set. To ensure that the trained potential is good enough for larger-scale MD simulations, we used active learning,[56,61] during MD, on a 1×2×3 supercell to determine if MTP is able to 'extrapolate' beyond the training set. Specifically, we used an extrapolation threshold of 2 to determine whether the energy and force predictions on the encountered structures are accurate or not. Once the extrapolation threshold of any encountered structure reached a pre-defined break value of 10, the MD simulation was stopped. Subsequently, the non-accurate structures were calculated with DFT, added to the training set, and the MTP was retrained. Notably, we found MTP to break during active learning only in MD simulations at 1200 K, indicating the highly disordered nature of the structures often encountered at this temperature, while the active learning runs at lower temperatures (900 K, 600 K, 300 K) did not break for 4000 ps. We generated a total of 569 structures during active learning that were added to the training dataset (i.e., 569 out of 3348 sampled for the optimized MTP) apart from our original AIMD simulations. Finally, we performed the large-scale and long-time MD simulations, without active learning, utilizing the retrained MTP after ensuring that the active learning runs on the 1×2×3 supercell were not broken for 4000 ps at 1200 K.

## 2.3. DFT and MD calculations

All our AIMD simulations were done at the DFT-level of theory using the Vienna ab initio simulation package,[71,72] by employing projector-augmented wave potentials.[73] Using a plane-wave kinetic energy cutoff of 520 eV, we employed the Hubbard $U$ corrected strongly constrained and appropriately normed (i.e., SCAN+$U$) functional to describe the electronic exchange and correlation.[74–77] The $U$ value applied on V $d$ orbitals was 1.0 eV, as derived in previous work.[74,75] We sampled the irreducible Brillouin zone using Γ-centred Monkhorst-Pack[78] $k$-point meshes with a density of at least 32 per Å (i.e., a minimum of 32 subdivisions along a unit reciprocal space vector), and we integrated the Fermi surface with a Gaussian smearing of width 0.05 eV. We did not preserve any underlying symmetry of any structure during our calculations. For relaxing the initial $V_2O_5$ structure, we converged both the total energies and atomic forces to within 0.01 meV and 30 meV/Å, respectively.

We performed AIMD simulations on a 1×2×3 supercell of $V_2O_5$ (and $MgV_2O_5$), corresponding to 84 (96) atoms, with a time step of 2 fs to ensure both accuracy and computational efficiency. We did classical MD simulations, also with a time step of 2 fs, based on our constructed MTPs, using the large-scale atomic/molecular massively parallel simulator (LAMMPS[79]) for two different supercell sizes of both $V_2O_5$ and $MgV_2O_5$, namely, 1×2×3 and 2×4×6. All AIMD and classical MD simulations used an *NVT* ensemble using a Nose-Hoover thermostat[80] and the velocity-Verlet algorithm[81,82] for integrating the equations of motion. For MTP-based MD simulations, we equilibrated the amorphous structures at each temperature using the *NVE* ensemble for 50 ps to randomize the velocities, followed by sampling (to estimate diffusivities) using the *NVT* ensemble for 4 ns. For calculating the intercalation voltage (see below) using MTP at 0 K, we performed a structural relaxation with LAMMPS by applying strict convergence criteria of $10^{-8}$ eV for the total energy and $10^{-8}$ eV/Å for the atomic forces.

For calculating average intercalation voltages, we obtained amorphous $MgV_2O_5$ structures, after a simulation time of 20 ps at different temperatures, both from AIMD and MTP-based MD, and ran a single self-consistent field (SCF) calculation with DFT, where the total energies were converged to within 0.01 meV. Subsequently, we removed the Mg atoms from each considered amorphous $MgV_2O_5$ structure and performed a SCF calculation on the corresponding deintercalated $V_2O_5$ structure, mimicking a topotactic Mg deintercalation process. We performed the SCF calculations on the AIMD/MD structures primarily to ensure that our energy scales are the same, especially with respect to the calculated total energy of Mg metal, which is required to set the reference in our voltage calculations to $Mg^{2+}/Mg$. Thus, our voltage calculations primarily capture the effect of disorder in the $V_2O_5$ structure. Given the SCF-calculated energies of $MgV_2O_5$ ($E_{MgV_2O_5}$) and $V_2O_5$ ($E_{V_2O_5}$), and the SCAN-calculated total energy of the hexagonal close-packed ground state of Mg metal ($E_{Mg}$), we (approximately) calculate the average Mg intercalation voltage, versus Mg metal, as in **Equation 1**.[83,84] The factor of two in the denominator of **Equation 1** corresponds to an exchange of two electrons per $Mg^{2+}$ exchanged, and $F$ is the Faraday's constant.

$$V = -\frac{E_{MgV_2O_5} - E_{V_2O_5} - E_{Mg}}{2F} \tag{1}$$

### 2.4. Diffusivity estimations

Ionic motion in solids is often measured using a tracer species with a diffusion coefficient, $D^*$, which is referred to as the self or tracer diffusivity. In principle, $D^*$ can be computed using a linear fit of the mean-squared displacement (MSD) of moving ions over time ($t$), as in **Equation 2**.[85] Note that the MSD in **Equation 2** is averaged over all the $N$ hopping ions, while $d$ indicates the dimensionality of the system (typically 3 in solids).

$$D^* = \frac{1}{2d} \lim_{t \to \infty} \frac{1}{N} \sum_{i=1}^{N} \frac{|\vec{r}_i(t) - \vec{r}_i(0)|^2}{t} \tag{2}$$

In AIMD and MD simulations, the total simulation time ($t_{tot}$) is often limited due to computational constraints, leading to statistical noise in the extracted MSD. Hence, we used the total mean squared displacement (TMSD, see **Equation 3**) that is averaged over multiple time intervals to mitigate statistical fluctuations, as proposed by He et al.[85] $N_{\Delta t}$ in **Equation 3** represents the number of possible time intervals with duration $\Delta t$.

$$\text{TMSD}(\Delta t) = \sum_{i=1}^{N} \frac{1}{N_{\Delta t}} \sum_{t=0}^{t_{tot}-\Delta t} |\vec{r}_i(t + \Delta t) - \vec{r}_i(t)|^2 \tag{3}$$

Subsequently, $D^*$ can be calculated as the slope of the MSD over time interval $\Delta t$ (see **Equation 4**), where $\text{MSD}(\Delta t) = \frac{\text{TMSD}(\Delta t)}{N}$ and varies linearly with $\Delta t$.

$$D^* = \frac{\text{MSD}(\Delta t)}{2d\Delta t} \tag{4}$$

For our MD simulations that were conducted over a total duration of 4000 ps, we chose $\Delta t = 200$ ps for calculating the MSD and $D^*$, while for AIMD simulations that lasted for 50 ps, we chose $\Delta t = 10$ ps.

Note that the TMSD and MSD as defined in **Equations 2** and **4** track the displacements of individual ions, while similar TMSD and MSD values can be calculated for the displacement of the centre of mass of the mobile ions – MSCD or mean squared centre-of-mass displacement – with $\Delta t$. The slope of MSCD($\Delta t$) versus $\Delta t$ yields the jump diffusivity ($D_J$) of the mobile ions, which in turn is related to the chemical diffusivity ($D_c$), as defined in Fick's first law,[86] by the thermodynamic factor ($\Theta$), as $D_c = \Theta D_J$.[85] The difference between $D_J$ and $D^*$ indicates the extent of cross-correlation among migrating ions in a system, which is often quantified using the Haven's ratio ($H_R$), as $H_R = \frac{D^*}{D_J}$. If a system exhibits completely uncorrelated motion (i.e., purely random motion), then $H_R = 1$, with deviations away from 1 indicating the extent of cross-correlation.

Finally, the diffusivity ($D$, tracer or jump) can be correlated with $E_a$ that controls the ionic migration via the Arrhenius relation of **Equation 5**. $D_0$ is the pre-exponential factor, which includes factors such as jump distance and attempt frequency, $k_B$ is the Boltzmann constant, and $T$ is the temperature. Thus, $E_a$, which is a material-specific property, can be obtained as the slope of the logarithm of $D$ vs. $1/T$. We utilised the DiffusionAnalyzer class of pymatgen[87,88] for post-processing our AIMD and MD calculations, extracting $D^*$ and $D_J$, and calculating $E_a$ based on $D^*$.

$$D = D_0 \exp\left(-\frac{E_a}{k_B T}\right) \qquad (5)$$

## 3. Results

### 3.1. Optimized MTP for MgV$_2$O$_5$ and V$_2$O$_5$

The parity plots between the final, optimized MTP and AIMD calculated energies and forces, in both the training and test sets are shown in **Figure S1.** Importantly, the constructed MTP demonstrates promising accuracy on the training data, with RMSE (MAE) on the energies and atomic forces at 4.68 (2.97) meV/atom and 0.243 (0.113) eV/Å, respectively. The RMSE (MAE) values on the test set are quite similar to the training set, namely, 3.16 (2.33) meV/atom and 0.244 (0.116) eV/Å on the energies and atomic forces, respectively, indicating MTP's ability to capture the interatomic interactions well and extrapolate beyond the training data. In terms of stress, our optimized MTP exhibits RMSE (MAE) of 0.65 (0.17) GPa and 0.50 (0.15) GPa on the train and test sets, respectively. To further verify the accuracy of our constructed MTP, we also examined the potential energy landscapes generated by MTP during MD and compared that with AIMD generated data across different temperatures (see **Figure S2**). For example, we observe a high degree of overlap between the potential energies calculated by MTP-based MD with AIMD values, beyond simulation times of 1 ps, for both $V_2O_5$ and $MgV_2O_5$ at the different temperatures considered, indicating the high degree of accuracy we can obtain with MTP-based MD. Thus, we believe that our optimized potential has the ability to describe the dynamics of both $V_2O_5$ and $MgV_2O_5$ systems with a high degree of accuracy.

### 3.2. RDFs and LRO

We examine the RDFs and LRO of the generated $V_2O_5$ and $MgV_2O_5$ structures to verify their amorphous nature upon melt-quench simulations. **Figure 2** presents the RDFs for $V_2O_5$ (panels a, b, and c) and $MgV_2O_5$ (d, e, and f), at select temperatures and calculated by both AIMD and MTP, while the calculated RDFs at other temperatures are compiled in **Figures S3**, **S4**, and **S5**. For example, **Figure 2a** displays the RDFs of the O-O (blue), V-V (red), and V-O (green) bonds in pristine $V_2O_5$, upon DFT relaxation at 0 K. Thus, **Figure 2a** represents the RDF in crystalline $V_2O_5$, which is characterized by sharp peaks for all types of bonds, and serves as a reference for comparing RDFs generated at other temperatures. **Figure 2b** shows the RDF in the AIMD generated $V_2O_5$, after 4 ps of simulation time, at 2000 K. In comparison to 0 K, the RDFs of all bonds show less-intense and broader peaks, with no discernable peaks beyond a distance of 6 Å, highlighting the significantly disordered state of $V_2O_5$ at 2000 K. Importantly, even at a high degree of disorder, the V-O RDF peak at ~2 Å is dominant, indicating that several of the V-O bonds that form the $VO_6$ octahedra or $VO_5$ square pyramids in crystalline-$V_2O_5$ are intact, implying that the disorder is mainly due to lack of connectivity among the V-O polyhedra (thus reducing LRO). Additionally, the RDFs predicted by MTP-based MD at 2000 K after 50 ps simulation time (**Figure 2c**) is quite similar to AIMD calculations, signifying that MTP captures the disordering of $V_2O_5$ accurately at 2000 K. Also, upon increasing the supercell size to 2×4×6 and at 2000 K in our MTP-based MD simulations, we can verify the complete melting (or amorphization) of the $V_2O_5$ structure, as characterized by the broad shoulder on all types of bonds from ~6.5 Å to ~16 Å (see **Figure S5b**).

In the case of $MgV_2O_5$, since we inserted Mg atoms into AIMD-generated $V_2O_5$ structures at different temperatures, we do not have a perfect crystalline representation of $MgV_2O_5$ within our dataset as compared to $V_2O_5$. Nevertheless, the AIMD-generated $MgV_2O_5$ structure, after 4 ps simulation time, at 300 K (**Figure 3d**) represents the closest structure to crystalline $MgV_2O_5$ that we have in our dataset, as characterized by its sharp peaks for all types of bonds, including Mg-O (orange), Mg-V (black), and Mg-Mg (purple). Analogous to our observation in $V_2O_5$, the disorder in $MgV_2O_5$ increases at higher temperatures, characterized by weaker peaks and broader shoulders beyond ~6.5 Å in the AIMD-generated structure at 1200 K (after 4 ps simulation time, **Figure 3e**). Also, the MTP-generated $MgV_2O_5$ structure at 1200 K (after 50 ps simulation time, **Figure 3f**) is quite similar to the AIMD-generated version, with the main differences arising out of Mg-Mg peak intensities at ~3.5 and 4 Å. Thus, our calculated RDFs indicate that we are able to disorder the $V_2O_5$ and $MgV_2O_5$ structure significantly during our AIMD simulations, which is captured equally well by our MTP-based MD simulations as well.

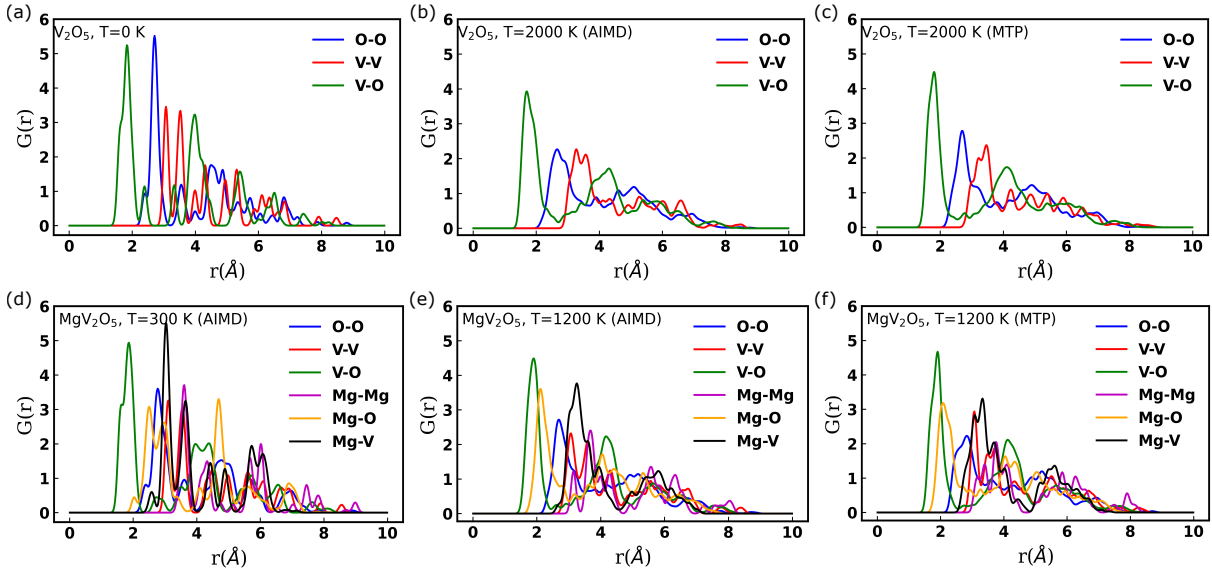

**Figure 2.** Radial distribution functions (RDFs or $G(r)$) of $V_2O_5$ (panels a-c) and $MgV_2O_5$ (panels d-f) calculated using AIMD or MTP under different temperatures. (a) Crystalline $V_2O_5$ (DFT-relaxed at 0 K), (b) AIMD-generated $V_2O_5$ at 2000 K after 4 ps, (c) MTP-generated $V_2O_5$ at 2000 K after 50 ps, (d) AIMD-generated $MgV_2O_5$ at 300 K after 4 ps, (e) AIMD-generated $MgV_2O_5$ at 1200 K after 4 ps, and (e) MTP-generated $MgV_2O_5$ at 1200 K after 50 ps. Blue, red, green, purple, orange, and black lines signify O-O, V-V, V-O, Mg-Mg, Mg-O, and Mg-V bonds, respectively.

To quantify the LRO in amorphous $MgV_2O_5$, at various temperatures, we used the open-source python package PyLRO,[89] which calculates the degree of directional disorder based on deviations in atomic spacings along different crystallographic directions. Specifically, we took structures based on our 1×2×3 supercell after a MTP-MD simulation time of 20 ps at 300 K, 600 K, 900 K, and 1200 K, calculated the extent of LRO along various directions, and plotted them in **Figure 3**. Note that the isosurfaces of **Figure 3** are depicted on a 3D Miller sphere, with different directions (and their associated Miller indices) indicated as text notations (along each axis). The maximum/minimum disorder in the structures and the disorder along the *a* [100], *b* [010], and *c* [001] axes for all structures are quantified in **Table S2**. We have also included visualizations of the structures used for calculating LRO in **Figure S6**.

Importantly, we observe an intuitive increase in structural disorder (or decrease in LRO) with increasing temperature, as given by the maximum (minimum) disorders of 0.11 (0.01), 0.12 (0.02), 0.21 (0.05), and 0.34 (0.07) at 300 K, 600 K, 900 K, and 1200 K, respectively (see **Table S2** and **Figure 3**). While there is some degree of similarity in the extent of (dis)order in the structures at 300 K and 600 K, there is a sharp increase in disorder at 900 K and beyond. At all temperatures, we observe an intermediate level of (dis)order, i.e., in between the maximum and minimum values, along the [100], [010], and [001] axes, indicating that the direction that breaks the LRO the most (or least) is not along the Cartesian basis vectors. Overall, we observe a progressive decline in the LRO of MTP-generated amorphous $MgV_2O_5$ with increasing temperature, which indicates the robustness of our melt-quench process and the reliability of our constructed MTP.

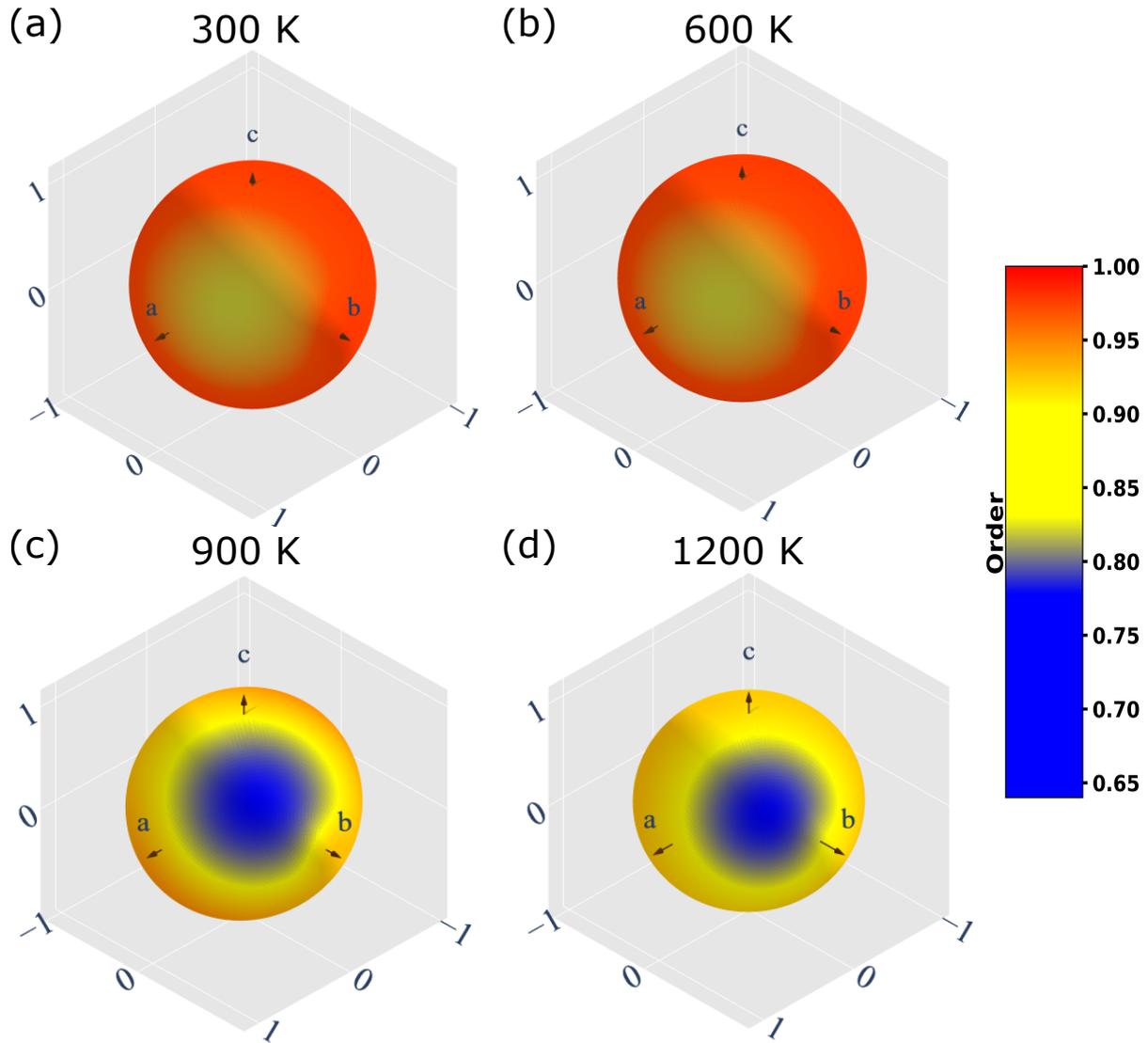

**Figure 3.** Isosurfaces of calculated LRO, plotted on 3D Miller spheres, in MTP-generated MgV$_2$O$_5$ structures at (a) 300 K, (b) 600 K, (c) 900 K, and (d) 1200 K, after a simulation time of 20 ps. Cartesian axes of the MgV$_2$O$_5$ structure are indicated as text annotations on the spheres. Blue (red) regions of an isosurface indicate low (high) LRO.

### 3.3. Voltage predictions

To quantify the impact of amorphization on the voltages of Mg intercalation into V$_2$O$_5$, we calculate the average voltages (versus Mg metal) based on AIMD/DFT (blue bars) and MTP (red) structures generated at 600 K, 900 K, and 1200 K, and compare the values to the average voltage obtained for the crystalline V$_2$O$_5$ structure (denoted as 0 K) in **Figure 4**. For calculating the AIMD voltage in the crystalline structure, we performed a full DFT relaxation of crystalline V$_2$O$_5$ and crystalline MgV$_2$O$_5$ structures, as available in the ICSD. In the case of MTP, we performed similar structure relaxations of both crystalline structures using LAMMPS.

Subsequently, we did a single SCF calculation of the MTP-relaxed structure using the SCAN+$U$ functional. Note that performing the SCF calculations captures the effect of structural disorder that is induced at different temperatures rather than the effect of the temperature itself on the calculated voltages, i.e., the temperatures in **Figure 4** are a proxy to indicate the extent of disorder in (Mg)V$_2$O$_5$ structures used in our voltage calculations.

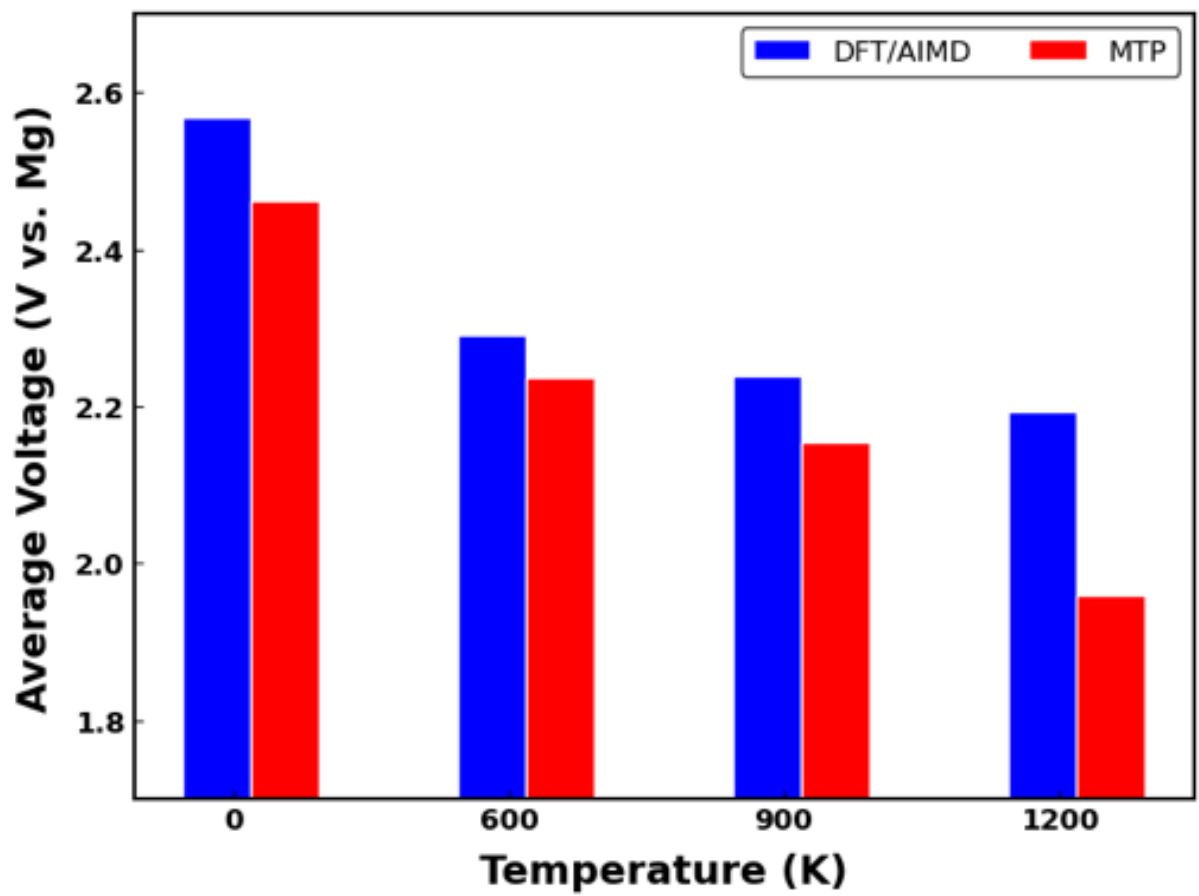

**Figure 4.** Comparison of average (de)intercalation voltage of Mg in V$_2$O$_5$ versus Mg metal, as calculated by AIMD/DFT (blue bars) and MTP (red bars), at different temperatures. 0 K indicates calculations done on the crystalline structure of V$_2$O$_5$ and MgV$_2$O$_5$.

The SCAN+$U$ calculated voltage for crystalline V$_2$O$_5$ (~2.56 V) is in agreement with the calculated (~2.52 V) and experimental (~2.3 V) voltage reported in literature.[22,27] Importantly, the AIMD calculated voltages decrease with increasing disorder in the structure, as indicated by the temperatures at which the structures are generated (see **Figure 3**). For example, the AIMD average voltage drops to ~2.3 V, ~2.23 V, and ~2.19 V at 600 K, 900 K, and 1200 K, respectively, reflecting a drop of ~0.26-0.37 V (10.2-14.4%) compared to the crystalline structure with increasing disorder. The drop in average voltage is reflective of the fact that the disruption in LRO affects the local bonding environment of Mg as it is intercalated into the structure, i.e., the potential energy surface at the Mg intercalation sites in the amorphous structure is not as deep as those encountered in the crystalline structure. Such a 'flattening' of the potential energy surface can indeed enhance Mg diffusion in the amorphous structure (see **Section 3.4**). Note that the drop in voltage in amorphous-$V_2O_5$ is only ~10-14% compared to crystalline-$V_2O_5$, signifying that the loss in energy density can be traded-off for a potential gain in Mg diffusivity within the material. Moreover, the average voltages in amorphous-$V_2O_5$ are higher than the SOTA chalcogenide Mg cathodes.[20,21,90] Additionally, the MTP-calculated voltage is in fair agreement with the first principles voltages for the crystalline structure (~2.46 V) as well as the disordered structures, with deviations of ~0.1 V at 600 K and 900 K, reflecting that the constructed MTP is able to describe the energetics fairly well. We do observe a significant underestimation in the MTP-based voltage at 1200 K (deviation of ~0.23 V vs. AIMD), which may be due to the specific structure that is sampled for the voltage calculation in the MTP simulation. Nevertheless, trends in our calculated voltages do indicate a possible drop of ~10% in amorphous $V_2O_5$ compared to its crystalline version, which needs to be accounted for if the amorphous structure is used as a cathode material in Mg batteries.

## 3.4. Mg diffusivities

**Figure 5a** displays the overall MSD($\Delta t$) as a function of $\Delta t$ (**Equation 4**), from MTP-based MD calculations on a 2×4×6 $MgV_2O_5$ supercell for 4000 ps. The yellow, blue, green, pink, and red lines in **Figure 5a** correspond to MSD data at 1200 K, 1000 K, 900 K, 600 K, and 300 K, and we observe linear relationships between MSD and $\Delta t$ for all temperatures. Note that the MSD displayed in **Figure 5a** is the overall MSD of Mg-ions, i.e., combining individual displacements along axes $a$, $b$, and $c$. The MSDs along individual axes are shown in **Figures S7** and **S8**. The actual MSD($t$) data from the MTP calculations as a function of simulation time ($t$, **Equation 2**) is shown in **Figures S9** and **S8**, which expectedly shows noisier statistics than the MSD($\Delta t$) data. Given the high computational costs associated with AIMD simulations, we only performed AIMD in 1×2×3 supercell, with the data compiled in **Figure S10**. Data from the corresponding MTP-MD simulations that we ran on a 1×2×3 supercell over 4000 ps is displayed in **Figures S11**, **S12**, and **S13**, with our MTP simulations on the 2×4×6 supercell displaying less noisy statistics. The linear relationship between log $D$ (with $D$ in units of $cm^2/s$) and $1/T$, as obtained from the MTP-based MD data on the 2×4×6 supercell at different temperatures is displayed in **Figure 5b**, with similar relationships derived from the 1×2×3 supercell AIMD and MTP data compiled in **Figures S10** and **S13**.

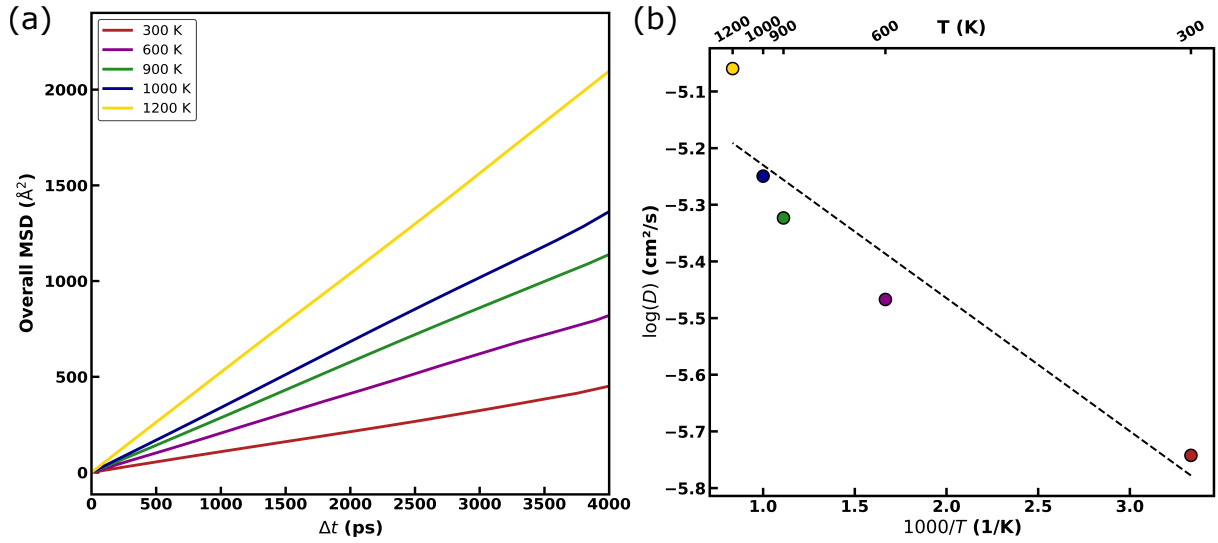

**Figure 5.** (a) Overall MSD(Δ$t$) values as a function of Δ$t$, calculated with MTP-MD over 4000 ps in a 2×4×6 MgV$_2$O$_5$ supercell. Yellow, blue, green, pink, and red lines represent calculated data at 1200 K, 1000 K, 900 K, 600 K, and 300 K, respectively. (b) Arrhenius plot of log $D$ (with $D$ in cm$^2$/s) versus 1000/$T$, derived from the data in panel a. Colors of the dots correspond to the temperatures in panel a. Dashed black line represents a linear fit.

Importantly, we estimate the MTP-calculated Mg-diffusivities in amorphous V$_2$O$_5$, based on the data in **Figure 5a**, to be 1.81×10$^{-6}$ cm$^2$/s at 300 K, 3.41×10$^{-6}$ cm$^2$/s at 600 K, 4.75×10$^{-6}$ cm$^2$/s at 900 K, 5.63×10$^{-6}$ cm$^2$/s at 1000 K, and 8.71×10$^{-6}$ cm$^2$/s at 1200 K. The associated $E_a$ with this variation of $D$ with $T$ is 47 meV. Such diffusivities and $E_a$ values are a remarkable improvement in Mg mobility in the amorphous V$_2$O$_5$ structure compared to the crystalline version. For example, previous studies[17,27,91] have estimated an $E_a$ of 600-750 meV and 975-1120 meV in the $\delta$ and $\alpha$ polymorphs of V$_2$O$_5$, respectively, corresponding to Mg diffusivities in the order of 10$^{-13}$ to 10$^{-16}$ cm$^2$/s in $\delta$ and 10$^{-20}$ to 10$^{-22}$ cm$^2$/s in $\alpha$ at 300 K. In comparison, amorphous V$_2$O$_5$ exhibits a Mg diffusivity in the order of 10$^{-6}$ cm$^2$/s at 300 K, which is a minimum improvement of about seven orders of magnitude in Mg diffusivity compared to the crystalline structure. Given that we have not changed the chemical composition here, this increase in diffusivity is entirely due to the amorphization of V$_2$O$_5$, resulting in a flatter potential energy surface.

Previously, Mg$^{2+}$ self-diffusivities have been reported to be in the range of 10$^{-11}$ to 10$^{-12}$ cm$^2$/s at 298 K in the thiospinel Mg$_x$Ti$_2$S$_4$ at $x$=0.35, with higher diffusivities (~10$^{-8}$ cm$^2$/s) reported at $x$~0 at 333 K.[8,21] Given that the thiospinel represents one of the SOTA Mg-cathode materials, we expect higher diffusivities in our amorphous oxide, by ~five and ~two orders of magnitude compared to the reported diffusivities in the thiospinel at 298 K and 333 K, respectively, representing a potentially significant gain in power performance. Moreover, amorphous V$_2$O$_5$ can yield a higher average voltage (by at least ~0.8 V) than Mg$_x$Ti$_2$S$_4$,[21,90] indicating a potential improvement in energy density as well. Thus, improvements in Mg mobility with the disruption of LRO provides a crucial handle that can enable the use of oxide (and other high energy density) cathodes for Mg batteries.

## 3.5. Degree of correlation

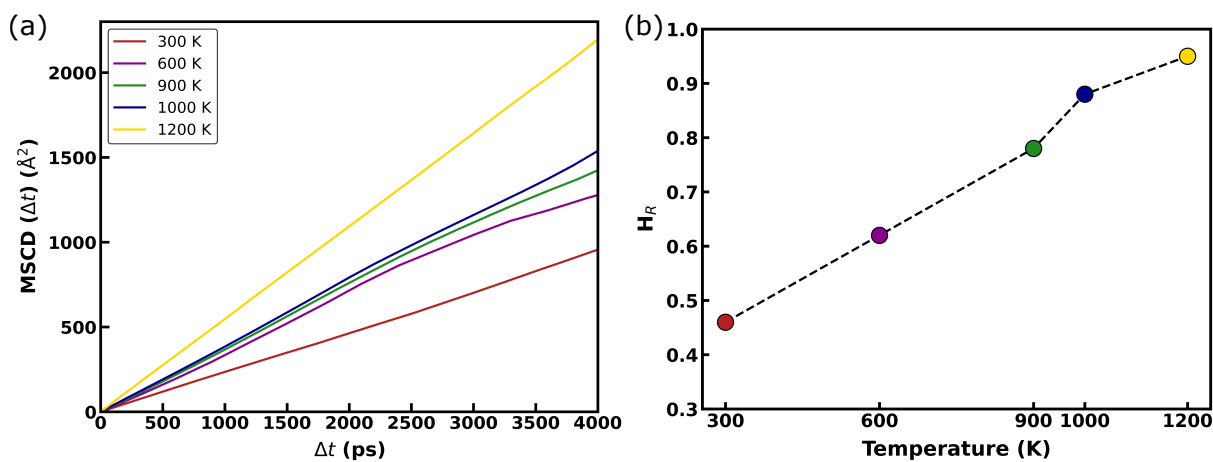

**Figure 6.** (a) Overall MSCD($\Delta t$) as a function of $\Delta t$ for amorphous MgV$_2$O$_5$ at different temperatures, as obtained from MTP-MD simulations in a 2×4×6 supercell over 4000 ps. (b) Variation of calculated $H_R$ with temperature. Yellow, blue, green, pink, and red colors in both panels represent data at 1200 K, 1000 K, 900 K, 600 K, and 300 K, respectively.

**Figure 6** plots the overall MSCD($\Delta t$) as a function of $\Delta t$ for amorphous-MgV$_2$O$_5$ (panel a) at different temperatures and the variation of $H_R$ with temperature (panel b) based on our MTP-MD calculations on the 2×4×6 supercell over 4000 ps. Calculated MSCD and $H_R$ data with MTP on the 1×2×3 supercell is provided in **Figure S14**. Notably, the MSCD's variation with $\Delta t$ demonstrates non-linearity at high $\Delta t$ at intermediate temperatures of 600 K and 900 K, which is expected since tracking centre-of-mass displacements can lead to noisier data than tracking individual atoms.[85] Importantly, our calculated $H_R$ shows a strong monotonic dependence of temperature in amorphous-MgV$_2$O$_5$, with values increasing from ~0.46 at 300 K to ~0.92 at 1200 K. Notably, diffusion along grain boundaries in metals (i.e., highly defective regions) is also known to be highly correlated, especially at low temperatures,[42] in qualitative agreement with our predictions for an amorphous oxide. At higher temperatures, the large availability of thermal energy ensures that atoms migrate in a more random (non-correlated) manner compared to lower temperatures, where concerted motion (cross-correlation) seems to occur due to stronger electrostatic interactions. Such concerted motion at low temperatures can possibly be characterized using experimental techniques and can provide a handle to further optimize the diffusivity of Mg ions within the amorphous $V_2O_5$ lattice by facilitating migration channels that allow for concerted Mg motion.

## 4. Discussion

In this work, we used a combination of AIMD and MLIP-MD simulations to explore amorphous $V_2O_5$ as a cathode material for Mg batteries, which form an important alternative technology in the space of beyond-Li-ion batteries. Specifically, we used melt-quench AIMD simulations and active learning (**Figure 1**) to generate ~3700 configurations of amorphous $V_2O_5$ and $MgV_2O_5$ to train accurate MTPs that yielded RMSE (MAE) of 3.16 (2.33) meV/atom and 0.244 (0.116) eV/Å on the test set energies and forces, respectively. We verified the amorphous nature of $V_2O_5$ and $MgV_2O_5$ structures obtained using both AIMD and MTP-MD, using RDF and LRO calculations (**Figures 2** and **3**), and noted the similarities in the RDFs obtained using AIMD and MTP-MD. Subsequently, we estimated the average Mg intercalation voltages, the Mg diffusivities (and associated $E_a$), and quantified the extent of cross-correlation in Mg motion, as a function of temperature using MTP-MD performed on large 2×4×6 supercells over 4 ns. We found that amorphization of $V_2O_5$ can result in a 10-14% drop in average Mg intercalation voltages compared to crystalline $V_2O_5$ (**Figure 4**), depending on the temperature. Importantly, we observed remarkably high Mg diffusivities (~$10^{-6}$ cm$^2$/s, **Figure 5**) in amorphous $MgV_2O_5$ at five different temperatures, with a low resultant $E_a$ of ~47 meV. Our calculated diffusivities in amorphous $MgV_2O_5$ are ~seven and ~five orders of magnitude higher compared to crystalline $Mg_xV_2O_5$ and thiospinel $Mg_xTi_2S_4$, respectively, highlighting the impact of amorphization on Mg mobility. Also, we found Mg diffusivities to be significantly cross-correlated at low temperatures ($H_R$~0.46 at 300 K, **Figure 6**) with the Mg motion becoming progressively random at high temperatures ($H_R$~0.92 at 1200 K). Overall, we observe amorphization to be a key handle that can significantly enhance Mg motion in oxides at room temperature, such as $V_2O_5$, thus enabling the use of high energy density cathodes with reasonable power performance, potentially resulting in the practical deployment of Mg batteries.

Our dataset consists of a total of 3725 configurations comprising $V_2O_5$ and $MgV_2O_5$ compositions, generated predominantly using melt-quench AIMD simulations, which was split 90:10 to create the train and test subsets, with our active learning workflow contributing ~569 configurations for optimizing our MTP. While our dataset is comprehensive and manages to capture the disordering of $V_2O_5$ quite well (**Figure 2**), it is limited by the range of temperatures and lack of intermediate compositions being sampled. Thus, expanding the range of temperatures being sampled (e.g., at every 100 K from 300 K to 1200 K) and the compositions (e.g., in steps of $\Delta x$=0.1 in $Mg_xV_2O_5$) will certainly improve the versatility and transferability of our constructed MTPs. Additionally, we used crystalline $V_2O_5$ ($\alpha$ polymorph) from ICSD as the starting structure for our melt-quench AIMD and added Mg within the amorphized-$V_2O_5$ structure instead of using $MgV_2O_5$ ($\delta$ polymorph) as a starting configuration, which adds some bias in our dataset. However, given that dataset generation carries the most computational expense in our study, it is impossible for us to encompass all relevant temperatures and compositions and remove all biases for training our MTP. Nevertheless, future works can build upon our dataset by incorporating more diverse data and generating MLIPs with better accuracy and transferability.

We chose MTP for our study for several reasons, including its computational speed using CPU-based processors, ability to learn swiftly from small datasets, and an integrated active learning framework.[56,57,61] Moreover, our previous work has demonstrated that MTP can generalize quite well across multi-component systems with diverse compositions.[60] Indeed, MTP performs quite well on our AIMD-based dataset, exhibiting low errors on train/test energies and forces, and exhibiting similar RDFs in both amorphous $V_2O_5$ and $MgV_2O_5$ structures compared to AIMD (**Figures 2**, **S3-S6**). However, MTP is an invariant potential and does not include features such as equivariance and message passing that is incorporated in more recent graph-based neural network potentials, such as neural equivariance interatomic potential[92] and multi atomic cluster expansion.[93] Thus, MTP is not the most general theoretical framework among MLIPs and does indeed exhibit limited flexibility, accuracy, and transferability.[94] Nevertheless, graph-based potentials are computationally expensive often requiring state-of-the-art GPU-based processors, are demanding on the memory, and are significantly slower on CPUs for MD simulations compared to MTP.[60] In any case, with the expansion of our dataset to more diverse temperatures and compositions, particularly on other oxide based systems, it may be worthwhile to construct graph-based potentials and/or fine-tune some of the available universal MLIPs.[95–98]

One limitation in our voltage predictions, particularly for structures extracted from higher temperatures, is that our voltage calculations have been done on a single snapshot of a structure instead of an average over an ensemble of possible configurations. Note that we don't necessarily expect any qualitative variations or significant quantitative differences between our calculations and ensemble averaged quantities, given that the predicted AIMD/MTP potential energies for both $V_2O_5$ and $MgV_2O_5$ (**Figure S2**) do not fluctuate significantly at different temperatures. Importantly, we observe a 10-14% drop in the average Mg intercalation voltage in amorphous $V_2O_5$ compared to its crystalline version (**Figure 4**), which can result in a 10-14% drop in energy density, a factor that needs to be accounted for in case amorphous $V_2O_5$ is used as a cathode in Mg batteries. To better understand and quantify the extent of voltage reductions due to amorphization of a given structure, similar studies on other oxides will be useful.

In terms of diffusivity estimates, sampling more temperatures (and compositions), longer simulations times, and larger supercells, can yield marginally better values than reported in our work (**Figure 5**). In any case, our MSD($\Delta t$) statistics as calculated by MTP-MD in our 2×4×6 supercell already appear quite robust (**Figures S7-S9**) and provides a good indication of the expected Mg mobility. More importantly, understanding the nature of correlated Mg motion (**Figure 6**) at lower temperatures will be quite challenging and important to further optimize the performance of amorphous $V_2O_5$. Specifically, the quenching rate and dopant additions can influence the short range order of amorphous $V_2O_5$, which in turn can impact the presence of 'open channels' for correlated Mg motion. While we provide a visual representation of Mg hops through a compiled video (provided as part of our GitHub repository, see 'Data and code availability' section), further characterization using electrochemical impedance spectroscopy, galvanostatic intermittent titration technique, and nuclear magnetic resonance can shed more insights into the underlying migration mechanism(s) that are active. Finally, note that although we report a fairly low $E_a$ (~47 meV, **Figure 5b**) for Mg motion in amorphous-$MgV_2O_5$, the definition of $E_a$ in amorphous systems is not as rigorous as in crystalline systems, since $E_a$ can vary significantly with temperature and the vacancy-based hopping mechanism may not be the only mechanism active in an amorphous structure. Nevertheless, our calculated $E_a$ provides an indication of the swift Mg mobility that is to be expected in amorphous $V_2O_5$ compared to its crystalline counterpart.

Our study demonstrates the potential of amorphous oxides, such as $V_2O_5$, as cathode materials for Mg batteries providing enhanced Mg mobility and a small drop in the intercalation voltage. Moving forward, similar investigations on other amorphous oxide and polyanionic chemistries, whose crystalline versions have shown promise as Li-ion cathodes,[99] such as Mn-, Co-, and Ni-based oxides, and Fe-, and Mn-based phosphates, can be carried out to explore their possible utility as Mg battery cathodes. The workflow established in this study (**Figure 1**) should provide a theoretical framework for executing studies on analogous cathode chemistries. Moreover, such studies will also provide statistics on the voltage drops, mobility enhancements, and (any) concerted migration mechanisms, which will be useful in improving the fundamental understanding of amorphous systems in general and in guiding the materials design for developing practical Mg batteries.

## 5. Conclusion

Magnesium batteries offer an alternative technological pathway with potentially higher volumetric energy density, lower costs, and better safety compared to the state-of-the-art LIBs. However, magnesium batteries require cathodes with high energy densities (e.g., oxides) to show reasonable Mg mobility (or power performance) to be practical. In this context, we used MD simulations powered by MLIPs (MTPs) that were trained on systematic melt-quench AIMD data to explore amorphous $Mg_xV_2O_5$ as a cathode for Mg batteries. Upon validating the MTP using active learning and verifying the amorphous nature of the generated structures via RDF and LRO calculations, we performed 4 ns simulations on a 2×4×6 supercell using MTP-based MD to estimate the Mg intercalation voltage and transport properties. Importantly, we observed a 10-14% drop in the average Mg intercalation voltage and a ~seven orders of magnitude improvement in Mg diffusivity (with a low $E_a$ of 47 meV) due to amorphization of the $V_2O_5$ framework. Our predicted Mg diffusivity in amorphous $V_2O_5$ is higher than in thiospinel $Mg_xTi_2S_4$, by ~five orders of magnitude at room temperature. Also, we observed significant cross-correlation in the Mg motion at room temperature, with the motion becoming progressively random at higher temperatures. Thus, we find amorphous $V_2O_5$ to be a promising cathode material for Mg batteries. More importantly, we expect amorphization of analogous oxides to be a key handle that can be used to design cathodes exhibiting high energy and reasonable power densities for Mg batteries. Finally, our theoretical workflow powered by MLIPs can be extended to explore a broader range of materials and properties, which will be a crucial step in accelerating the design and discovery of new amorphous materials for different applications.

**Author contributions**

V.C. performed all AIMD calculations, constructed MTP, and validated the potential using active learning. D.D. performed MTP-based MD simulations. V.C. wrote the initial draft of the paper. D.D. edited the draft and aided in the visualization of data. G.S.G. supervised all aspects of the work. All authors approve the final version of the manuscript.

**Acknowledgements**

G.S.G. acknowledges financial support from the Indian Institute of Science (IISc) Seed Grant, SG/MHRD/20/0020 and SR/MHRD/20/0013 and the Science and Engineering Research Board (SERB) of the Department of Science and Technology, Government of India, under sanction numbers SRG/2021/000201 and IPA/2021/000007. V.C. thanks the Institute of Eminence Post-doctoral fellowship awarded by the Indian Institute of Science for financial assistance. D.D. acknowledges the Indian Institute of Science for academic support and the Shell Fellowship for financial support. A portion of the density functional theory calculations showcased in this work were performed with the computational resources provided by the Supercomputer Education and Research Center, Indian Institute of Science. We acknowledge National Supercomputing Mission (NSM) for providing computing resources of 'PARAM Siddhi-AI', under National PARAM Supercomputing Facility (NPSF), C-DAC, Pune and supported by the Ministry of Electronics and Information Technology (MeitY) and Department of Science and Technology (DST), Government of India.

**Data and Code Availability**

All computed data, relevant scripts, and the best MLIPs constructed in this work are available freely for all via our [GitHub](#) repository.